\documentclass[twocolumn,showpacs,prl,english,showkeys,10pt,aps,unsortedaddress,superscriptaddress]{revtex4-1}

\usepackage{graphicx,amsmath,amssymb}
\usepackage[english]{babel}
\usepackage{hyperref}

\begin{document}

\title{Phase transitions in disordered systems: the example of the random-field Ising model in four dimensions}
\author{Nikolaos G.~Fytas}
\affiliation{Applied Mathematics Research Centre, Coventry University, Coventry CV1 5FB, United Kingdom}

\author{V\'{i}ctor Mart\'{i}n-Mayor} 
\affiliation{Departamento de F\'isica
  T\'eorica I, Universidad Complutense, 28040 Madrid, Spain}
\affiliation{Instituto de Biocomputac\'ion y F\'isica de Sistemas Complejos
  (BIFI), 50009 Zaragoza, Spain}

\author{Marco Picco}
\affiliation{LPTHE (Unit\'e mixte de recherche du CNRS UMR 7589), Universit\'e
  Pierre et Marie Curie - Paris 6, 4 place Jussieu, 75252 Paris cedex 05,
  France}

\author{Nicolas Sourlas}
\affiliation{Laboratoire de Physique Th\'eorique de l'Ecole Normale
  Sup\'erieure (Unit{\'e} Mixte de Recherche du CNRS et de l'Ecole Normale
  Sup\'erieure, associ\'ee \`a l'Universit\'e Pierre et Marie Curie, PARIS VI)
  24 rue Lhomond, 75231 Paris CEDEX 05, France}

\date{\today}

\begin{abstract}
\noindent
By performing a high-statistics simulation of the $D=4$ random-field
Ising model at zero temperature for different shapes of the
random-field distribution, we show that the model is ruled by a single
universality class.  We compute to a high accuracy the complete set of
critical exponents for this class, including the correction-to-scaling
exponent. Our results indicate that in four dimensions: (i)
dimensional reduction as predicted by the perturbative renormalization
group does not hold and (ii) three independent critical exponents are
needed to described the transition.
\end{abstract}

\pacs{05.50.+q,75.10.Nr,02.60.Pn,75.50.Lk}

\maketitle

{\it Introduction} --- The random-field Ising model (RFIM)~\cite{imry:75} is
maybe the simplest disordered system in Physics~\cite{parisi:94}. Applications
in hard and soft condensed matter Physics are many (see
e.g.~\cite{nattermann:98,belanger:98,fytas:15b}), and their numbers
increase~\cite{franz:13a,franz:13b,biroli:14}.  The RFIM Hamiltonian is
\begin{equation}
\label{H}
{\cal H} = - J \sum_{<xy>} S_x S_y - \sum_{x} h_x S_x \; ,
\end{equation}
with the spins $S_x = \pm 1$ occupying the nodes of a hyper-cubic lattice in
space dimension $D$ with nearest-neighbor ferromagnetic interactions and $h_x$
independent random magnetic fields with zero mean and dispersion $\sigma$.  

The Renormalization Group (RG) suggests that $D$ is an all-important variable
(no less than temperature $T$)~\cite{wilson:74}. Indeed, at low temperature
$T$ and for small-enough disorder (i.e., $\sigma\ll J$), we encounter the
ferromagnetic phase, provided that $D \geq 3$ \cite{imbrie:84,bricmont:87}. A
phase transition to a disordered, paramagnetic phase occurs upon increasing
$T$ or $\sigma$. Yet, for $D=2$, the tiniest $\sigma>0$ suffices to destroy
the ferromagnetic phase~\cite{aizenman:90}.  Furthermore, perturbative RG
(PRG) computations, employing the mathematically unorthodox replica trick to
restore the translation invariance broken by disorder~\cite{edwards:75}, tell
us that the upper critical dimension is $D_\mathrm{u}=6$~\cite{aharony:78}
(Mean Field is quantitatively accurate if $D>D_\mathrm{u}$).

The RFIM and branched polymers are unique among disordered systems: a
supersymmetry \cite{parisi:79c} makes it possible to analyze the PRG to all
orders of perturbation theory~\cite{parisi:81}. Supersymmetry predicts
dimensional reduction: the RFIM critical behaviour in dimension $D$ would be
the same of a non-disordered ferromagnet in dimension
$D-2$~\cite{young:77,parisi:79c}. Yet, see above, the RFIM orders in $D=3$
while the ferromagnet in $D=1$ does not. 

The failure of the
PRG begs the question: Is there
an intermediate dimension $D_{\mathrm{int}}<D_\mathrm{u}$ such that the PRG is
accurate for $D>D_{\mathrm{int}}$? The issue is obviously relevant to all
disordered systems~\footnote{See Refs.~\cite{parisi:94,mezard:92,mezard:94,parisi:02} for
  possible sources of non-perturbative behavior.}.

Yet, the RFIM is a peculiar disordered system. The relevant RG fixed-point
is believed to lie at $T=0$~\cite{villain:85,bray:85b,fisher:86b}. Therefore, in order to
describe the critical behavior one needs \emph{three} independent critical
exponents and \emph{two} correlation functions, namely the connected and
disconnected propagators, $C^{\mathrm{(con)}}_{xy}$ and
$C^{\mathrm{(dis)}}_{xy}$ \footnote{For $T>0$,
  $C^{\mathrm{(con)}}_{xy}=\overline{\langle S_x S_y\rangle - \langle
    S_x\rangle\langle S_y\rangle}/T\,$, hence the name connected
  propagator.}. At the critical point and for large $r$ ($r$: distance between
$x$ and $y$), they decay as
\begin{equation}\label{eq:anomalous}
C^{\mathrm{(con)}}_{xy}\!\equiv\!\frac{\partial\overline{\langle
  S_x\rangle}}{\partial h_y} \!\sim\!
\frac{1}{r^{D-2+\eta}}\,;\
 C^{\mathrm{(dis)}}_{xy}\!\equiv\!
\overline{\langle S_x\rangle\langle S_y\rangle}\! \sim\!
\frac{1}{r^{D-4+\bar\eta}}\,,
\end{equation}
where the $\langle \ldots \rangle$ are thermal mean values as computed for a
given realization, a \emph{sample}, of the random fields $\{h_x\}$. Over-line
refers to the average over the samples. The relationship between the anomalous
dimensions $\eta$ and $\bar\eta$ is hotly debated, and it is one of our main
themes here, as it entails the correct parametrization of the
neutron-scattering line-shape~\cite{slanic:99,ye:04}.  Supersymmetry predicts
$\eta=\bar\eta$.

We also recall phenomenological scaling as an alternative to the
PRG~\cite{imry:75,villain:85, bray:85b, fisher:86b}.  The prediction $\bar\eta
= 2 \eta$ by Schwartz and coworkers~\cite{schwartz:86,schwartz:91,gofman:93},
although not a consequence of phenomenological scaling, has gained ground
thoughout the years. However Tarjus and
coworkers~\cite{tissier:11,tissier:12,tarjus:13} have suggested that rare
events, neglected in~\cite{schwartz:86,schwartz:91,gofman:93}, spontaneously
break supersymmetry at the intermediate dimension $D_{\mathrm{int}}\approx
5.1$. For $D>D_{\mathrm{int}}$ replica predictions hold: supersymmetry is
valid and $ \overline{\eta} = \eta $. For $ D < D_{\mathrm{int}} $, instead,
there are three independent critical exponents.

Unfortunately, both the perturbative and the phenomenological RG approaches
lack predictions allowing for detailed comparisons with experiments. In this
context numerical simulations become a crucial tool. This is especially true
at $T=0$, where fast polynomial algorithms~\cite{auriac:85,goldberg:88} allow
us to find exact ground states for a wide range of accessible system sizes
$L$. This approach has been used mainly at
$D=3$~\cite{ogielski:86,auriac:97,swift:97,sourlas:99,hartmann:99,middleton:01,hartmann:01,middleton:02,dukovski:03,wu:05,wu:06,fytas:13,picco:14}
but also for higher dimensions on a smaller
scale~\cite{swift:97,hartmann:02,middleton:02b,ahrens:11}, although having a
strong command over the $D$-dependency of the random-field criticality would
be desirable and is the motivation of the current work.
 
Noteworthy, claims of universality violations for the RFIM at $D\geq 3$ have
been quite frequent when comparing different distributions of random
fields~\cite{auriac:97,swift:97,sourlas:99,hartmann:99}.  Fortunately, using
new techniques of statistical analysis~\cite{fytas:15b}, it has been possible
to show that, at least in $D=3$, these apparent universality violations are
merely finite-size corrections to the leading scaling
behavior~\cite{fytas:13,picco:15}. We also note the numerical bound
$2\eta-\bar\eta \leq 0.0026(10)$~\cite{fytas:13} which is valid in
$D=3$~\footnote{On the other hand, $0\leq 2\eta-\bar\eta$ is valid for all
  $D$~\cite{gofman:93}}.

Here, we report the results of large-scale zero-temperature numerical
simulations at $D=4$. Our state-of-the-art analysis~\cite{fytas:15b,fytas:13}
provides high-accuracy estimates for the critical exponents $\eta$,
$\bar\eta$, and $\nu$, as well as for other RG-invariants, indicating that
dimensional reduction does not hold at this particular dimensionality. A clear
case for universality is made by comparing Gaussian-and Poissonian-distributed
random fields, but only after taming the strong scaling corrections. Finally,
we present overwhelming numerical evidence in favor of $2\eta-\bar\eta>0$,
indicating that three independent critical exponents are needed to describe
the transition and, furthermore, that the intermediate space dimension where
supersymmetry gets restored is larger than four.

{\it Simulation details and finite-size scaling} ---
We consider the Hamiltonian~(\ref{H}) on a $D=4$ hyper-cubic
lattice with periodic boundary conditions and energy units 
$J=1$. Our random fields $h_{x}$ follow either
a Gaussian $({\mathcal P}_G)$, or a Poissonian $({\mathcal P}_P)$ distribution:
\begin{equation}
{\mathcal P}_G(h,\sigma) = {1\over \sqrt{2 \pi
      \sigma^2}} e^{- {h^2 \over 2\sigma^2}}\;,\ {\mathcal P}_P(h,\sigma) = {1\over 2 |\sigma| } e^{- {|h| \over
\sigma}}\;,
\end{equation}
where $-\infty< h < \infty$. For both distributions $\sigma$ is our single
control parameter.

We simulated lattice sizes from $L=4$ to $L=60$. For each $L$ and $\sigma$
value we computed ground states for $10^7$ samples, see the Supplemental
Material {\bf SM}~\cite{supplemental}. For comparison, 3200 samples of $L=32$ were simulated in~\cite{hartmann:02} and 5000 samples
of $L=64$ in ~\cite{middleton:02b}.

From simulations at a given $\sigma$,
we computed $\sigma$-derivatives and extrapolated to neighboring $\sigma$
values by means of a reweighting method~\cite{fytas:15b}. We computed the
second-moment correlation length~\cite{amit:05} for each of the two
propagators $C^{\mathrm{(con)}}$ and $C^{\mathrm{(dis)}}$ in
Eq.~\eqref{eq:anomalous}, $\xi^{\mathrm{(con)}}$ and $\xi^{\mathrm{(dis)}}$,
as well as the corresponding susceptibilities $\chi^{\mathrm{(con)}}$ and
$\chi^{\mathrm{(dis)}}$. We also computed the dimensionless Binder ratio $U_4
= \overline{\langle m ^4 \rangle}/\overline{\langle m^2\rangle}^2$ and the
ratio $U_{22}=\chi^{\mathrm{(dis)}}/[\chi^{\mathrm{(con)}}]^2$ that gives a
direct access to the difference of the anomalous dimensions
$2\eta-\bar\eta$. For additional technical details see Ref.~\cite{fytas:15b}.

\begin{figure}
\centerline{\includegraphics[scale=0.30,
    angle=0]{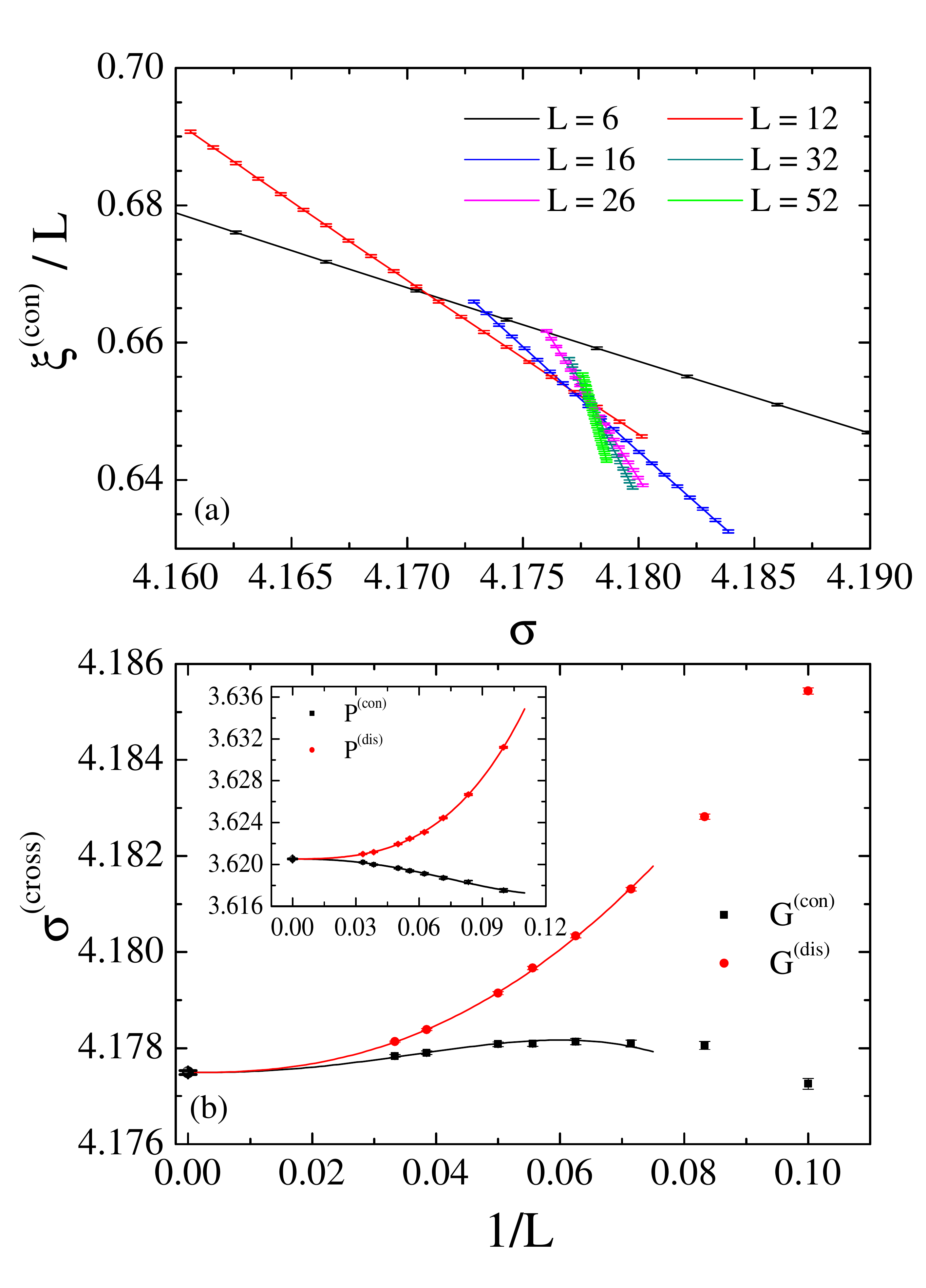}} \caption{(color online) {\bf Top:} Connected
  correlation length in units of the system size $L$ vs.  $\sigma$ (we show data only for some characteristic $L$
  values for clarity' sake).  Due to scale invariance, all curves should cross at the critical
  point $\sigma_\mathrm{c}$. Yet, small systems deviate from the large-$L$
  scale-invariant behavior. {\bf Bottom:} For Gaussian random fields, crossing
  points $\sigma_{\mathrm{c},L}$ of pair of lattice-sizes ($L,2L$) for
  $\xi^{\mathrm{(dis)}}/L$ and $\xi^{\mathrm{(con)}}/L$, as a function of
  $1/L$.  Lines are fits to Eq.~\eqref{QFS}, constrained to yield a common
  extrapolation to $L=\infty$ (depicted as a black circle at the origin in
  this figure and in the following ones). {\bf Inset:} Same as in bottom
  panel, but for the case now of Poissonian random fields. In all figures the
  notation G$^{\mathrm{(con),(dis)}}$ [or P$^{\mathrm{(con),(dis)}}$]
  distinguishes the type of crossing point (or the type of random fields,
  i.e., Gaussian or Poissonian).}
\label{F1CP}
\end{figure}

We followed the quotients-method approach to finite-size
scaling~\cite{amit:05,nightingale:76,ballesteros:96}. In this method,
one considers dimensionless quantities $g(\sigma,L)$ that, barring
correction to scaling, are $L$-independent at the critical point. We
consider two such $g$, namely $\xi^{\mathrm{(dis)}}/L$ and
$\xi^{\mathrm{(con)}}/L$ (also $U_4$ is dimensionless). Given a
dimensionless quantity $g$, we consider a pair of lattices sizes $L$
and $2L$ and determine the crossing $\sigma_{\mathrm{c},L}$, where
$g(\sigma_{\mathrm{c},L},L)= g(\sigma_{\mathrm{c},L},2L)$, see
Fig.~\ref{F1CP}--top. For each random-field distribution we compute two
such $\sigma_{\mathrm{c},L}$, one for $\xi^{\mathrm{(dis)}}/L$ and one
for $\xi^{\mathrm{(con)}}/L$. Crossings approach the critical point as
$\sigma_{\mathrm{c}}-\sigma_{\mathrm{c},L}={\cal
  O}(L^{-(\omega+1/\nu)})$, with $\omega$ being the leading
corrections-to-scaling exponent.

Dimensionful quantities $O$ scale with $\xi$ in the thermodynamic
limit as $\xi^{x_O/\nu}$, where $x_O$ is the scaling dimension of
$O$. At finite $L$, we consider the quotient $Q_{O,L} = O_{2L}/O_L$ at
the crossing (for dimensionless magnitudes $g$, we write
$g^\mathrm{cross}_{L}$ for either $g_{L}$ or $g_{2L}$, whichever show
less finite-size corrections)
\begin{equation}\label{eq:QO}
Q_{O,L}^\mathrm{cross} = 2^{ x_O/\nu} +
  O(L^{-\omega}) \; \; ; \; \; g^\mathrm{cross}_{L} = g^{\ast} +
  O(L^{-\omega})\,.
\end{equation}
$Q_{O}^\mathrm{cross}$ (or $g^\mathrm{cross}_{L}$) can be evaluated both at
the crossing for $\xi^{\mathrm{(dis)}}/L$ or $\xi^{\mathrm{(con)}}/L$. The two
choices differ only in the scaling corrections, an opportunity we shall
use. The RG tells us that $x_O$, $g^\ast$, $\omega$, and $\nu$, are
universal. We shall compute the critical exponents using Eq.~\eqref{eq:QO}
with the following dimensionful quantities: $\sigma$-derivatives [$x_{D_\sigma
    \xi^{\mathrm{(con)}}}= x_{D_\sigma \xi^{\mathrm{(dis)}}}=1+\nu$],
susceptibilities [$x_{\chi^{\mathrm{(con)}}}= \nu(2-\eta)$ and
  $x_{\chi^{\mathrm{(dis)}}}= \nu(4-\bar\eta)$] and the ratio $U_{22}$
[$x_{U_{22}}=\nu(2\eta-\bar\eta)$]. We also note the ambiguity with
$g^\mathrm{cross}_{L}$. If you study, say, $g=\xi^{\mathrm{(dis)}}/L$ at the
crossings of $\xi^{\mathrm{(con)}}/L$, you may focus just as well on $g_{2L}$,
or on $g_L$. Scaling corrections can be the smallest in either case. The
corrections-minimizing choices are $g^\mathrm{cross}_{L}=g_{2L}$ for
$\xi^{\mathrm{(dis)}}/L$, $g^\mathrm{cross}_{L}=g_{L}$ for
$\xi^{\mathrm{(con)}}/L$, and $g^\mathrm{cross}_{L}=g_{2L}$ for $U_4$.

Now, an important issue is evinced in Fig.~\ref{F1CP}--bottom: The
size evolution is non monotonic (for a spectacular example see
Fig.~\ref{FigU4}--{\bf SM}~\cite{supplemental}). In
other words, our accuracy is enough to resolve sub-leading corrections
to scaling.

We take into account sub-leading corrections in an effective way.
Let $X_L$ be either $g^\mathrm{cross}_{L}$ or the effective
scaling dimension $x^\mathrm{(eff)}_O/\nu = \log
Q_{O}^\mathrm{cross}(L)/\log 2$, recall Eq.~\eqref{eq:QO}. We
consider two different fits ($a_k,b_k,c_k,d_k$ for $k=1,2$ are
scaling amplitudes): \\ (i) The quadratic fit (QF) which is
\begin{eqnarray}
X_L&=& X^{\ast}+a_1 L^{-\omega}+ a_2 L^{-2\omega} \; ,\label{QF}\\
\sigma_{\mathrm{c},L}&=&\sigma_c+ b_1 L^{-(\omega+\frac{1}{\nu})}+
b_2 L^{-(2\omega+\frac{1}{\nu})} \;.\label{QFS}
\end{eqnarray}
(ii) However, $\omega$ turns out to be so large, that
$L^{-2\omega}$ terms (certainly present) are maybe not the most
relevant correction. Hence we consider also the leading + analytic
corrections fit [(L+A)F],
\begin{eqnarray}
X_L&=& X^{\ast}+c_1 L^{-\omega}+ c_2 L^{-(2-\eta)} \; ,\label{LF+A}\\
\sigma_{\mathrm{c},L}&=&\sigma_c+ d_1 L^{-(\omega+\frac{1}{\nu})}+
d_2 L^{-(2-\eta+\frac{1}{\nu})}\label{LF+AS} \;.
\end{eqnarray}
The $L^{-(2-\eta)}$ term is due to the non-divergent analytic
background. We plug $2-\eta \simeq 1.8$ in the (L+A)F.

Since both fits are well motivated only when $L$ is large enough,
we restrict ourselves to data with $L\geq L_\mathrm{min}$. To
determine an acceptable $L_\mathrm{min}$ we employ the standard
$\chi^2$-test for goodness of fit, where $\chi^2$ is computed using the
complete covariance matrix. In practice, we found that both types
of fit give compatible results. In the following, we present the
results of the QF [for the results of the (L+A)F, see
Tab.~\ref{table1}].

\begin{figure}
\centerline{\includegraphics[scale=0.30, angle=0]{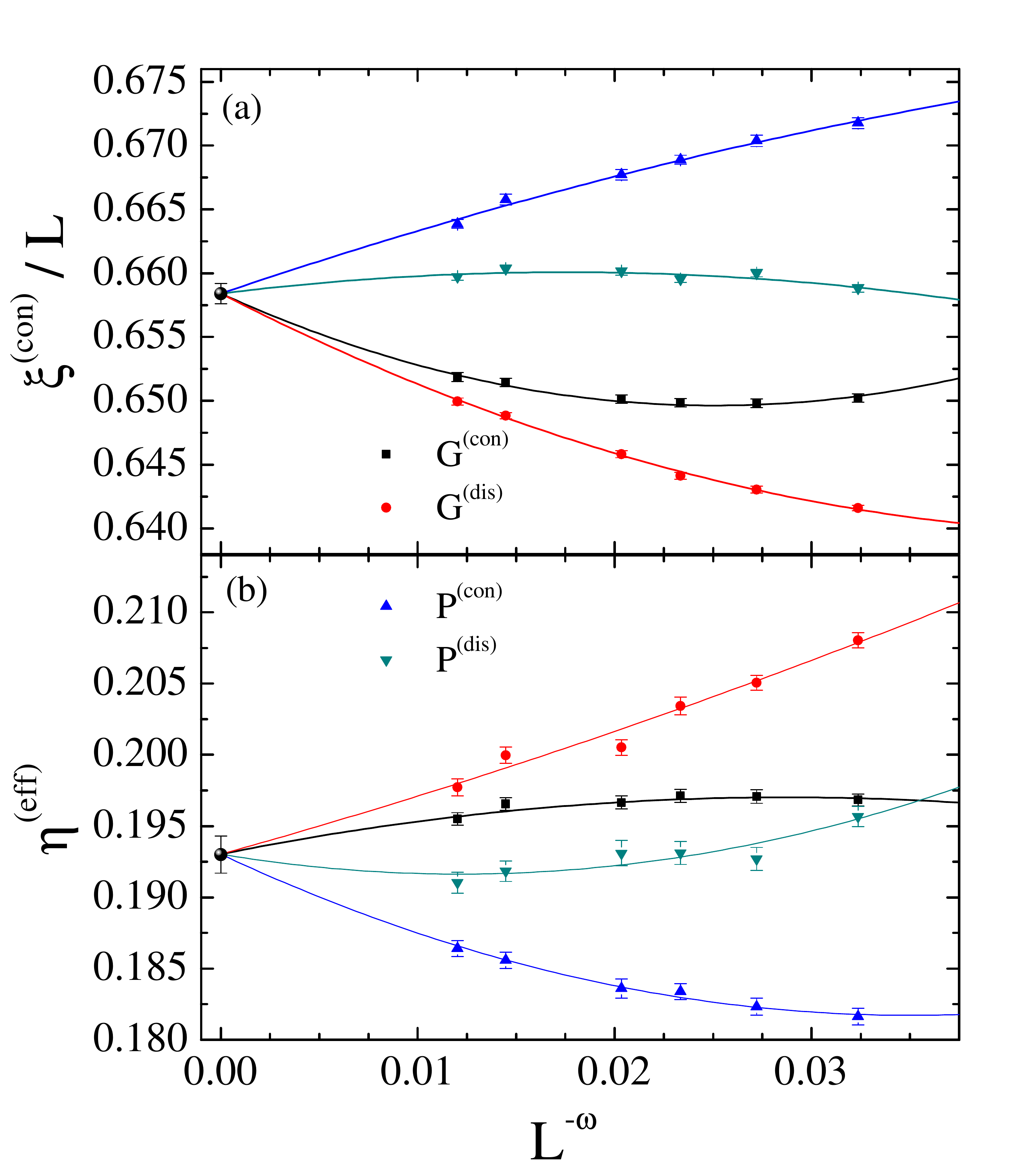}}
\caption{(color online) {\bf Top:} $\xi^{\mathrm{(con)}}/L$ vs. $L^{-\omega}$ at
the crossing points shown in the upper panel of Fig.~\ref{F1CP}.
{\bf Bottom:} The same as in top panel,
  but for $\eta^{\mathrm{(eff)}}$. Lines correspond to the joint fit reported in
  Table~\ref{table1}.} \label{F2xi}
\end{figure}

{\it Results} ---
The procedure we follow is standard by now~\cite{ballesteros:98}. 
The first step is the estimation of the
corrections-to-scaling exponent $\omega$. Take, for instance,
$\xi^{\mathrm{(con)}}/L$. For each pair of sizes $(L,2L)$ we have four
estimators, Fig.~\ref{F2xi}--top: two crossing points, either
$\xi^{\mathrm{(con)}}/L$ or $\xi^{\mathrm{(dis)}}/L$, and two disorder
distributions. Rather than four independent fits to Eq.~\eqref{QF}, we
perform a single joint fit: we minimize the combined $\chi^2$
goodness-of-fit, by imposing that the extrapolation to $L=\infty$,
$(\xi^{\mathrm{(con)}}/L)^{\ast}$, as well as exponent $\omega$ are
common for all four estimators (only the scaling amplitudes differ).
We judge from the final $\chi^2$ value whether or not the fit is fair.

Furthermore, one can perform joint fits for several magnitudes,
say $\xi^{\mathrm{(con)}}/L$ and $\eta$. Of course, the
extrapolation to $L=\infty$ is different for each magnitude, but a
common $\omega$ is imposed. However, when we increase the number
of magnitudes, the covariance matrix becomes close to singular due
to data correlation, and the fit becomes unstable. Therefore, we
limit ourselves to $\xi^{\mathrm{(con)}}/L$ and $\eta$, see
Fig.~\ref{F2xi} and Fig.~\ref{Figchi}--{\bf SM}~\cite{supplemental}.
We obtain a fair fit, Table~\ref{table1}, by considering pairs $(L,2L)$ with $L\geq L_{\mathrm{min}}=14$.

The rest of the quantities of interest are individually extrapolated,
following the same procedure, but now fixing $\omega=1.30(9)$ (For the
extrapolation of $\xi^{\mathrm{(dis)}}/L$ and $U_4$ see Figs.~\ref{F3xid}--{\bf SM} and
\ref{F4U4}--{\bf SM} in~\cite{supplemental}). In fact, the extrapolations in
Tab.~\ref{table1} have two error bars. The first error, obtained from the
corresponding joint fit to Eq.~\eqref{QF}, is of statistical origin. The
second error is systematic and takes into account how much the extrapolation
to $L=\infty$ changes in the range $1.21<\omega <1.39$.

\begin{figure}
\centerline{\includegraphics[scale=.30, angle=0]{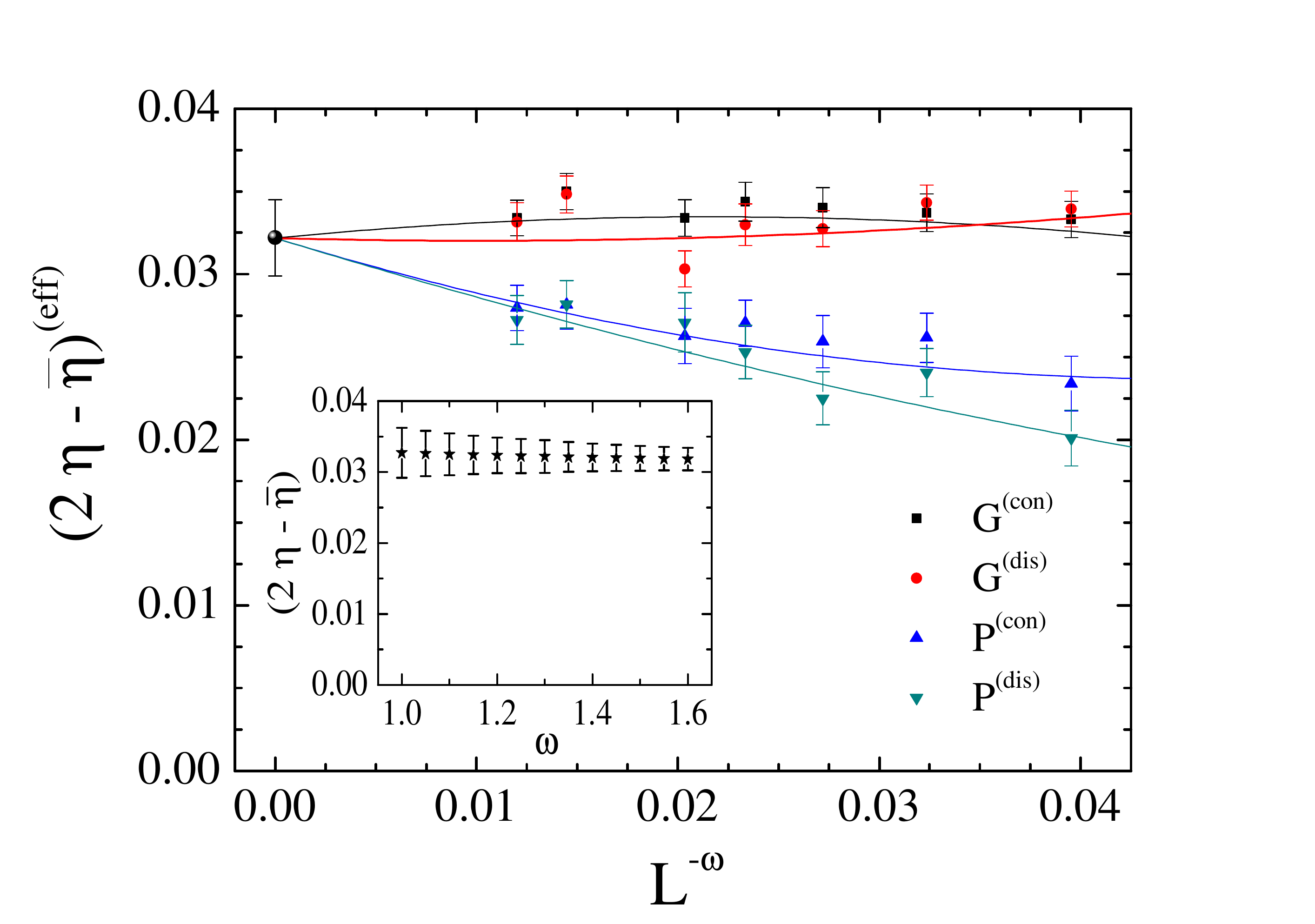}}
\caption{(color online) Effective anomalous dimension difference $2 \eta
-\bar\eta$ vs. $L^{-\omega}$ at the crossing points shown in
 the upper panel of Fig.~\ref{F1CP}. Lines correspond to a joint fit to
Eq.~\eqref{QF} with $\omega=1.3$. {\bf Inset:} The extrapolation
to large-$L$ of $2 \eta -\bar\eta$ is essentially
$\omega$-independent.} \label{F5U22}
\end{figure}

\begin{figure}
\centerline{\includegraphics[scale=0.30, angle=0]{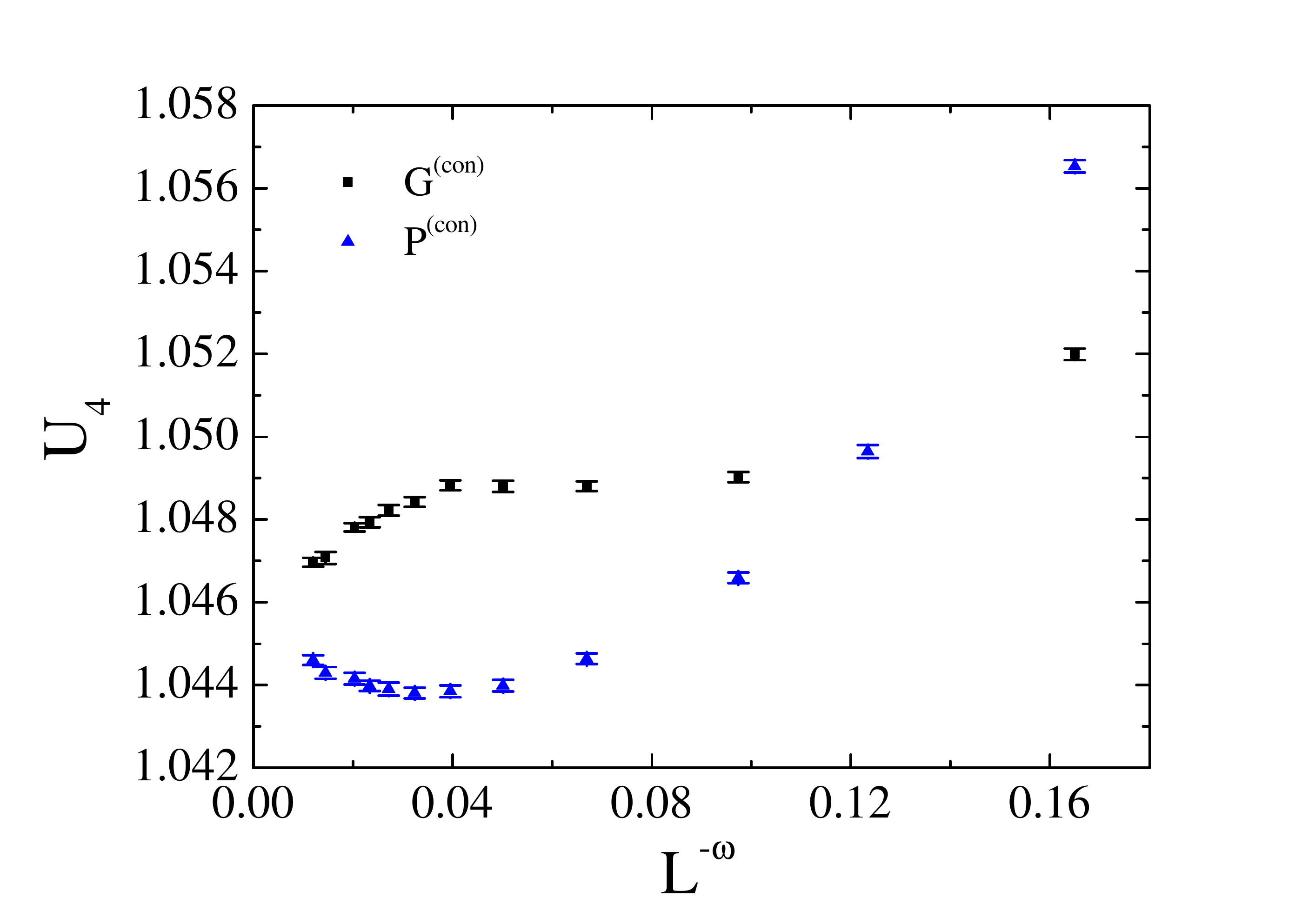}}
\caption{(color online) $U_{4}$ vs. $L^{-\omega}$ for the complete lattice-size
spectrum ($\omega=1.3$).} \label{FigU4}
\end{figure}

\begin{figure}
\centerline{\includegraphics[scale=0.30,
angle=0]{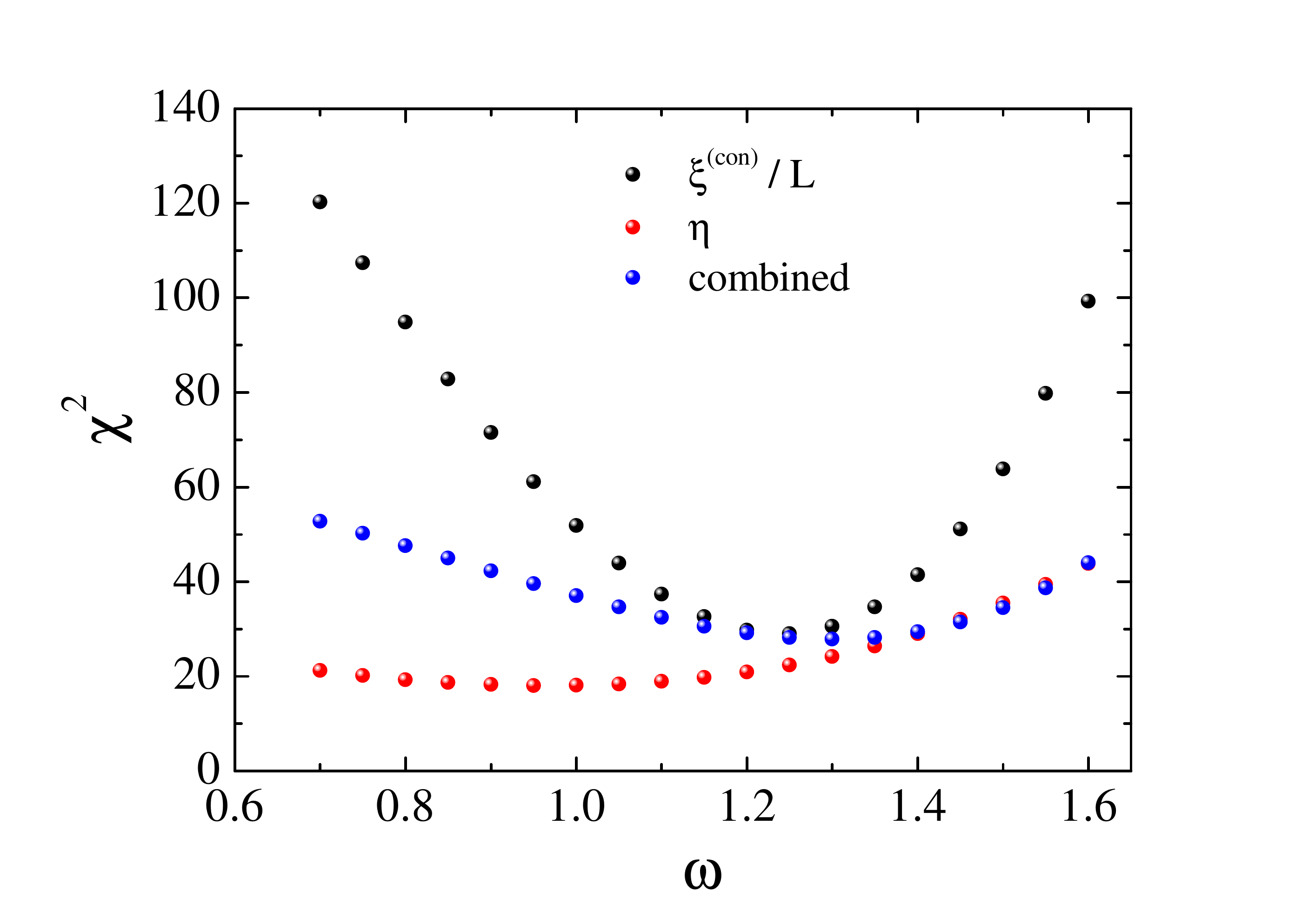}} \caption{(color online) Minimum of $\chi^{2}$ as a
function of $\omega$ for the fits shown in Fig.~2 of the main
manuscript, referring to the universal ratio $\xi^{\rm (con)}
/L$, the anomalous dimension $\eta$, and the combined data.}
\label{Figchi}
\end{figure}

Our main result is illustrated in Fig.~\ref{F5U22}, where we show
$\log U_{22}/\log 2$ which is a direct measurement of the
difference $2\eta -\bar\eta$. This extrapolation is particularly
easy, because $L_{\mathrm{min}} = 12$ is enough to obtain a good
fit and a value $2\eta - \bar\eta = 0.0322(24)$. Furthermore, the
dependency on $\omega$ of the large-$L$ extrapolation is weak, as
shown in Fig.~\ref{F5U22}--inset.  We conclude with high
confidence that $ 2 \eta -\bar\eta $ is different than zero, in
support of the three-exponent scaling scenario.

We also determined the effective exponent $\nu^{\mathrm{(eff)}}$ from the
$\sigma$-derivatives of $\xi^{\mathrm{(con)}}$ and $\xi^{\mathrm{(dis)}}$ (see
Fig.~\ref{Fignu}--{\bf SM}~\cite{supplemental}). The fits were acceptable even with
$L_{\mathrm{min}}=8$ (Table~\ref{table1}).

Previous less accurate numerical estimates for the Gaussian distribution of random
fields that did not take into account sub-leading
corrections are given by Hartmann: $\nu=0.78(10)$,
$\sigma_{\mathrm{c}}=4.18(1)$, $\eta=0.18(1)$, and $\bar\eta=0.37(5)$
(so that $2\eta-\bar\eta\approx -0.01$)~\cite{hartmann:02} and
Middleton: $\nu=0.82(6)$ and
$\sigma_{\mathrm{c}}=4.179(2)$~\cite{middleton:02b}.
We may also quote the functional RG estimates
$2\eta-\bar\eta=0.08(4)$, $\nu=0.81(3)$ and
$\eta=0.24(1)$~\cite{tissier:12}, close but incompatible with our
findings, probably due to the truncation of
the functional RG equations.

{\it Conclusions} --- We have carried out a zero-temperature numerical study
of the random-field Ising model in four dimensions. By using two types of the
random-field distribution and a proper finite-size scaling scheme we have been
able to show universality and to determine with high accuracy the three
independent critical exponents, $\eta$, $\bar\eta$, and $\nu$, that are needed
to describe the transition, as well as other renormalization group invariants.
We stress the non-trivial difference between the anomalous
dimensions $2\eta - \bar\eta = 0.0322(24)$ which is ten times larger than its
corresponding value at $D = 3$~\cite{fytas:13}. We thus provided decisive
evidence in favor of the three-exponent scaling scenario and the spontaneous
supersymmetry breaking~\cite{tissier:11,tissier:12} at some $D_{\mathrm{int}}
> 4$, against the (restricted) scaling
picture~\cite{schwartz:86,schwartz:91,gofman:93}.

Let us conclude by mentioning our preliminary simulations in five
dimensions, not reported here. Critical exponents in $D=5$ turn
out to be very close to those of the $D=3$ pure Ising
ferromagnet, as supersymmetry and dimensional reduction predict.
This finding suggests that $D_{\mathrm{int}}\approx 5$, in
quantitative agreement with Refs.~\cite{tissier:11,tissier:12}.
We intend to pursue this investigation in the near future. As for
the suspected upper critical dimension, $D_{\mathrm{u}}=6$,
characteristic logarithmic scaling violations have been
reported~\cite{ahrens:11}, but still await detailed confirmation.
These two final steps will give us access to the full picture of
the RFIM scaling behavior.

\begin{table}
\caption{Summary of results. The second column is the outcome of a fit to
  Eq.~\eqref{QF} while the fourth column is obtained fitting to
  Eq.~\eqref{LF+A} [yet, critical points $\sigma_\mathrm{c}$ were obtained
    from Eqs.~\eqref{QFS} or~\eqref{LF+AS}, correspondingly]. The first row
  reports a joint fit for $\omega$, $\xi^{\mathrm{(con)}}/L$ and $\eta$. The
  remaining quantities were individually extrapolated to $L=\infty$. $\chi^2$
  is the standard figure of merit (DOF: number of degrees of freedom in the
  fit). }\label{table1}
\begin{center}
\begin{tabular}{ | l || c | c | c | c |}
\hline
&     QF & $ \chi^2/{\mathrm{DOF}}$      &  (L + A)F &  $\chi^2/{\mathrm{DOF}}$\\
 \hline
$\omega $   &  1.30(9)       &     &  1.60(14)     &\\
$\xi^{\mathrm{(con)}}/L$      &  0.6584(8)   &   27.85/29   & 0.6579 (+6/-4) & 40.33/37\\
$\eta$           &  0.1930(13) &    & 0.1922(10)  &\\
 \hline\hline
$\sigma_{\mathrm{c}} (G)$   &  4.17749(4)(2)    &  5.6/7   & 4.17750(4)(2)  & 3.2/7\\\hline
$\sigma_{\mathrm{c}} (P)$  &  3.62052(3)(8)  & 8.85/11 &  3.62060(3)(1)     & 9.8/11\\\hline
\hline
$U_4$    &  1.04471(32)(14)     &   10/11 &  1.04490(36)(9)       &    8.57/11\\\hline
$\xi^{\mathrm{(dis)}} / L $  &  2.4276(36)(34)  &   16/15  &  2.4225(41)(20)   &   14/15\\
\hline
$\nu$             & 0.8718(58)(19)    & 62.9/55  & 0.8688(64)(11)     & 59.8/55\\\hline
$2 \eta -\bar\eta $  & 0.0322 (23)(1)   &16.0/19  &  0.0322(25)(1)     & 16.1/19\\
\hline
\end{tabular}
\end{center}
\end{table}

\begin{acknowledgments}
Our $L=52, 60$ lattices were simulated in the {\em MareNostrum}
and {\em Picasso} supercomputers (we thankfully acknowledge
the computer resources and assistance provided by
the staff at the {\em Red Espa\~nola de Supercomputaci\'on\/}).
N.G.F. was supported from  Royal Society
Research Grant  N$^o$ RG140201 and from a Research
Collaboration Fellowship Scheme of Coventry University. V.M.-M.
was supported by MINECO (Spain) through research contract
N$^o$ FIS2012-35719C02-01.
\end{acknowledgments}

\bibliographystyle{apsrev4-1}
\bibliography{biblio.bib}

\newpage

\textbf{Supplemental Material}

\vspace{0.3cm}

The algorithm used to generate the ground states of the system was
the push-relabel algorithm of Tarjan and
Goldberg~\cite{goldberg:88}. We prepared our own C version of the
algorithm, involving a modification proposed by Middleton \emph{et
al.}~\cite{middleton:01,middleton:02,middleton:02b} that removes
the source and sink nodes, reducing memory usage and also
clarifying the physical
connection~\cite{middleton:02,middleton:02b}. Additionally, the
computational efficiency of our algorithm has been increased via
the use of periodic global
updates~\cite{middleton:02,middleton:02b}.

We simulated systems with lattice sizes within the range $L=4 -
60$. In particular, we considered the sizes
$L=\{4,5,6,8,10,12$, $14,16,18,20,24,26,28,30,32,36,40,52,60\}$, so
that we created up to $12$ pairs of system sizes $(L, 2L)$ in
order to apply the quotients method. For each set ($L, \sigma$)
and for each field distribution, Gaussian and Poissonian, we
simulated $10^7$ independent random-field realizations. As we
applied the quotients method at both the crossings of the
connected and disconnected correlation length over the system
size, i.e., $\xi^{\rm (con)} / L$ and $\xi^{\rm (dis)} / L$,
typically the sets of simulations were doubled for each system
size as the crossings between the connected and disconnected cases
varied. Note also, that throughout the main manuscript we have
used the notation $\rm {Z}^{\rm (x)}$, where Z denotes the
distribution, i.e., G for Gaussian and P for Poissonian, and the
superscript x refers to the connected (con) and disconnected (dis)
type of the universal ratio $\xi^{\rm (x)} / L$.

We present in Fig.~\ref{FigU4} the result for
the Binder cumulant $U_4$ for the complete lattice-size spectrum,
starting from $L=4$ (compare to Fig.~\ref{F4U4} below where
results are shown for $L\geq 16$). Results are given for both the
Gaussian and Poissonian distributions estimated at the crossings
of the $\xi^{\rm (con)} / L$. Although $U_{4}$ converges toward a
unique value in the large size limit in support of universality,
we observe that by including data for smaller lattices that there
is a strong inflection. This justifies our choice of the inclusion
of further corrections-to-scaling for smaller system sizes and
also consists a clear illustration of misleading behavior at
different scaling regimes. We believe that this latter point was
behind the strong violations of universality claimed in previous
works of the RFIM.

In  Fig.~\ref{Figchi}, we show a complementary plot
that supports the claim of Fig.~2 of the main manuscript. In
particular, we plot the $\omega$-minimization attempt of the merit
of the fit shown Fig.~\ref{Figchi}, in terms of $\chi^{2}$, for both
$\xi^{\rm (con)} / L$ and $\eta$ separately and also for the
combined fit that has provided us with the final values of
$\xi^{\rm (con)} / L$, $\eta$, as well as the
corrections-to-scaling exponent $\omega$.

The extrapolation to the thermodynamic
limit of the critical $\xi^{\mathrm{(dis)}}/L$ is illustrated in
Fig.~\ref{F3xid} (a fair fit was obtained with
$L_\mathrm{min}=14$). Similarly, also $U_4$ converges to its
universal limit, as shown in Fig.~\ref{F4U4} ($L_{\mathrm{min}} =
16$ in this case).

Finally, in Fig.~\ref{Fignu} we illustrate the
infinite limit-size extrapolation of the effective exponent $\nu$
of the correlation length, for both types of distributions
studied. The solid lines are a joint polynomial fit of second
order in $L^{-\omega}$ including data points for $L\geq 8$,
extrapolating to $L^{-\omega} = 0$, as shown by the filled circle
in the figure. We remind the reader that for the effective
exponent $\nu$ we have two sets of data for each of the two
distributions coming from the connected and disconnected
correlation lengths.

\begin{figure}
\centerline{\includegraphics[scale=0.30, angle=0]{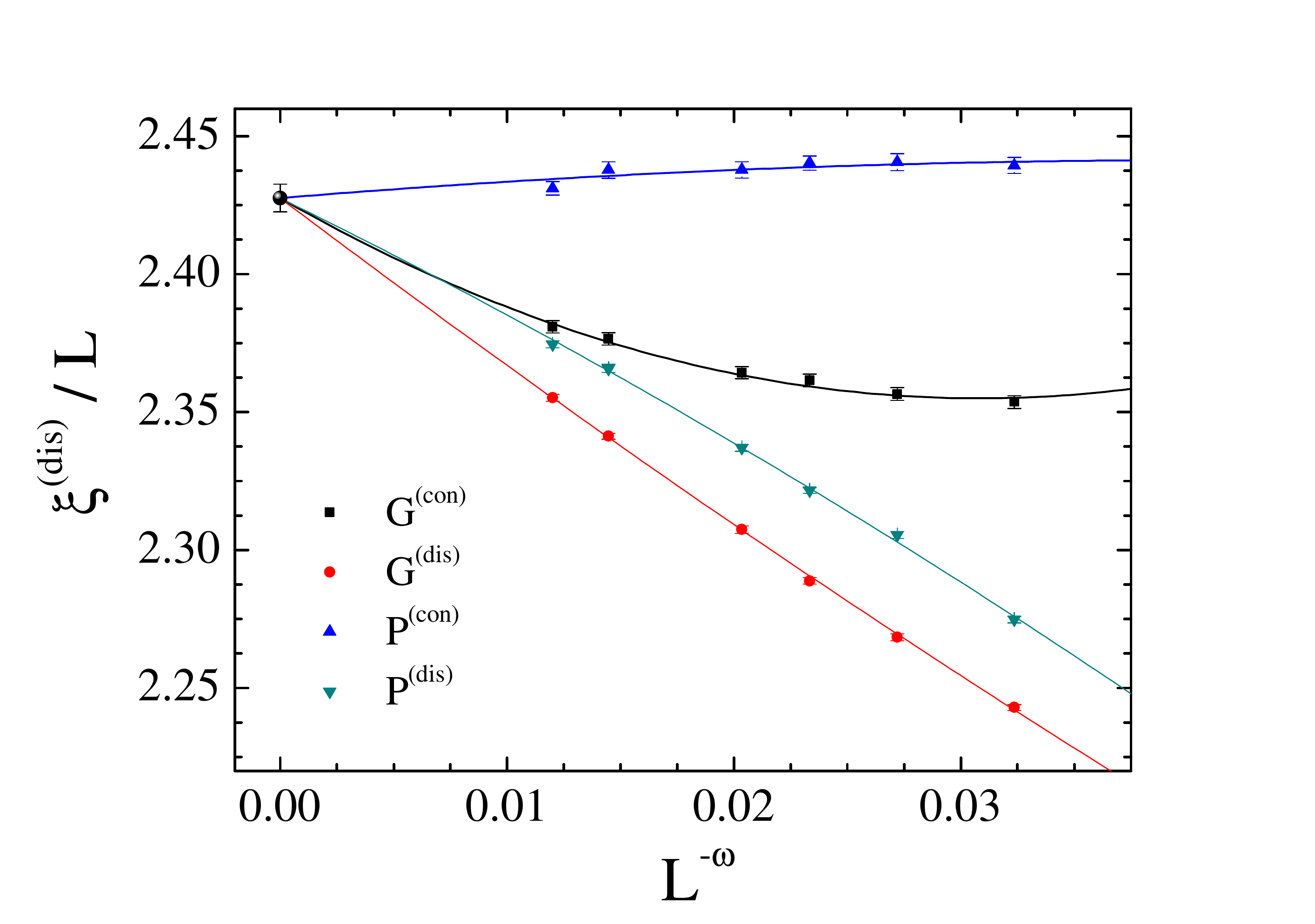}}
\caption{(color online) $\xi^{\mathrm{(dis)}}/L$ vs. $L^{-\omega}$ at the
crossing points shown in the upper panel of Fig.~1 in the main text. Lines
correspond to a joint fit to Eq.~(5) in the main text (with $\omega=1.3$).}
\label{F3xid}
\end{figure}

\begin{figure}
\centerline{\includegraphics[scale=.30, angle=0]{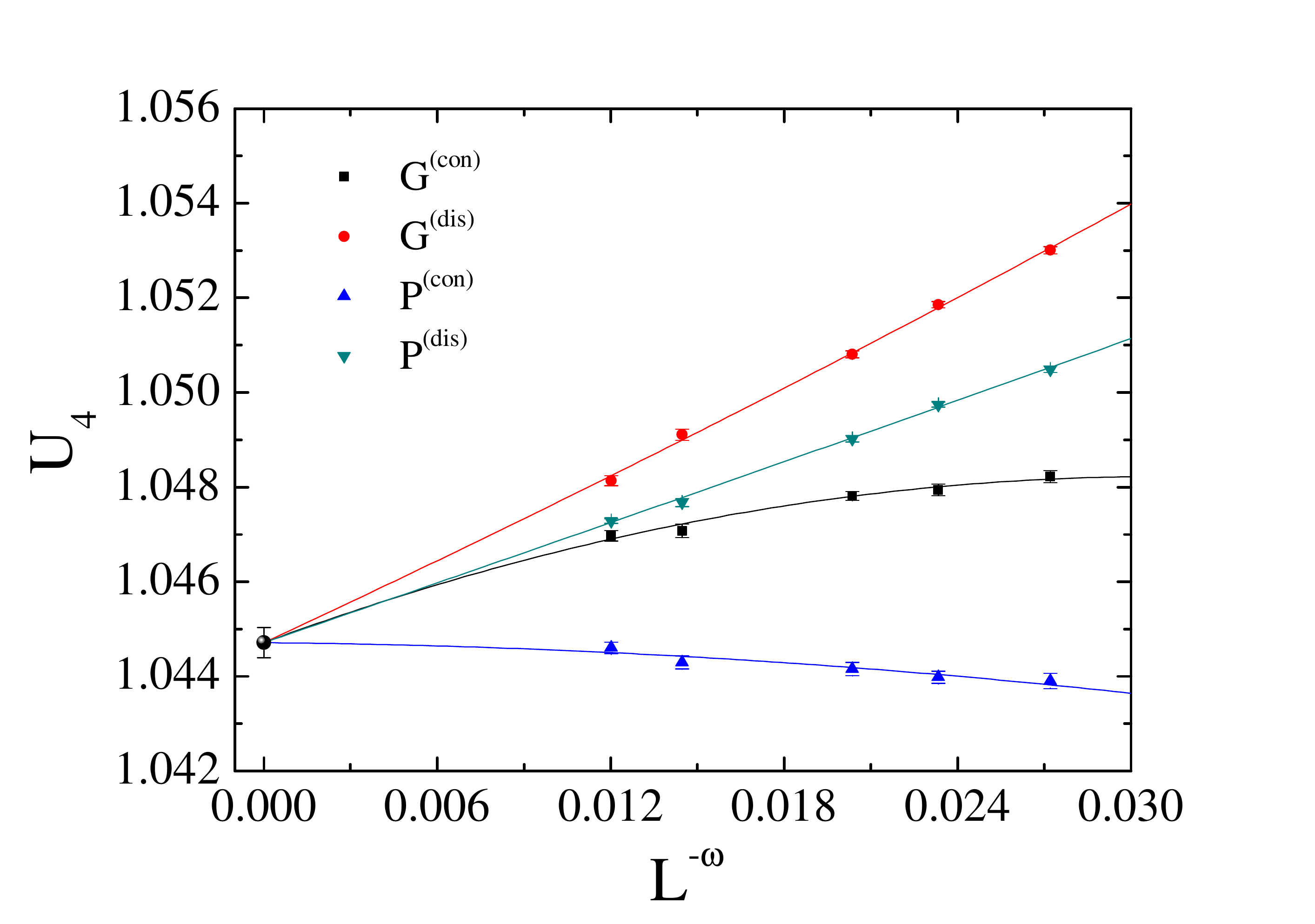}}
\caption{(color online) $U_4$ vs. $L^{-\omega}$ at the crossing points shown in
 the upper panel of Fig.~1 in the main text. Lines correspond to a joint fit to
Eq.~(5) in the main text (with $\omega=1.3$).}
\label{F4U4}
\end{figure}

\begin{figure}
\centerline{\includegraphics[scale=0.30, angle=0]{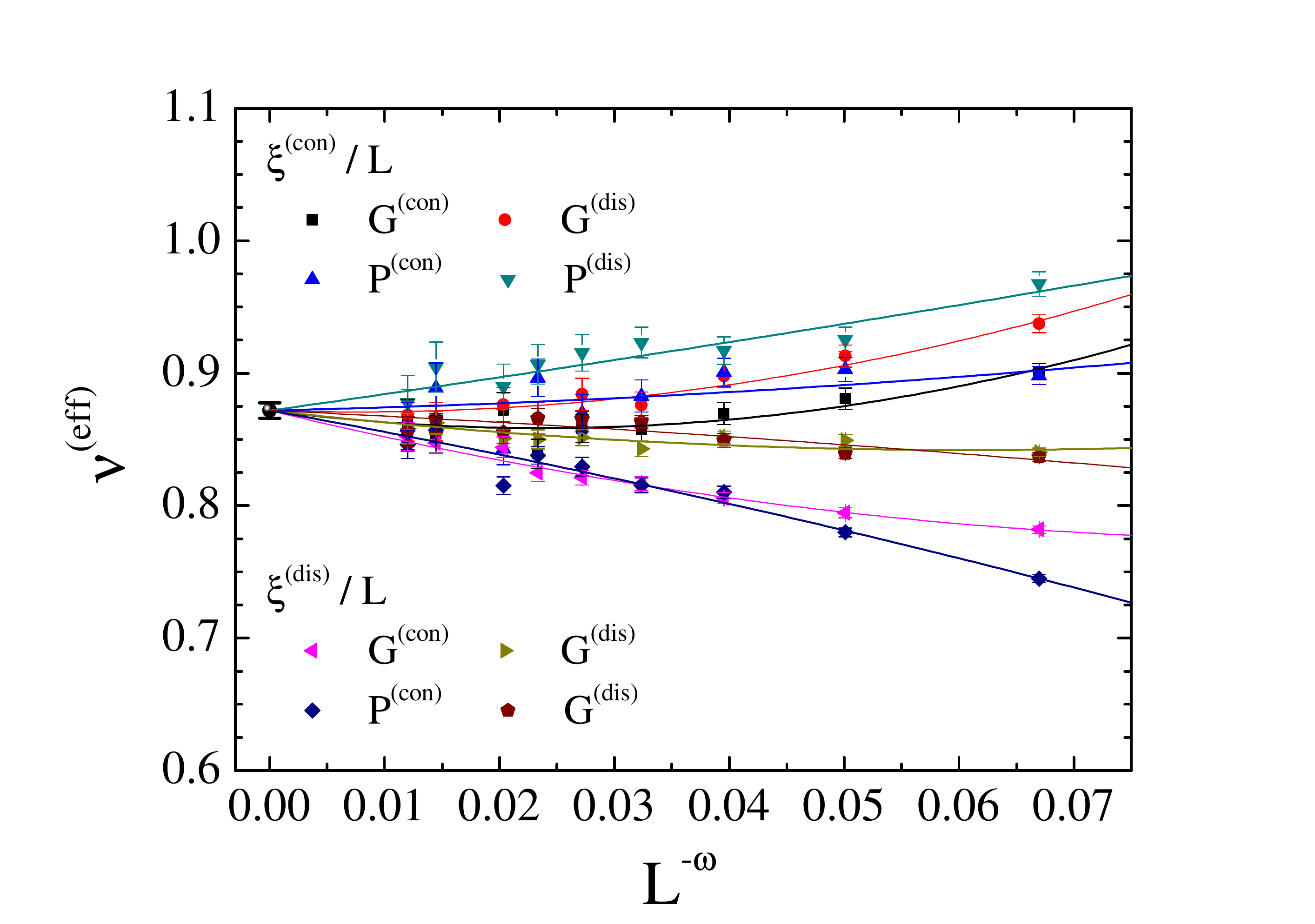}}
\caption{(color online) Infinite limit-size extrapolation of the effective
  critical exponent $\nu$. Lines correspond to a joint fit to Eq.~(5) in the
  main text (with $\omega=1.3$).} \label{Fignu}
\end{figure}

\end{document}